\documentclass[a4paper, 10pt, twocolumn]{article} 

\usepackage[utf8]{inputenc}
\usepackage[T1]{fontenc}

\usepackage[hmargin={2cm,2cm}, vmargin={2cm,2cm}]{geometry}
\usepackage{graphicx}
\usepackage{url,hyperref,lineno,microtype,subcaption}
\hypersetup{colorlinks=true,linkcolor=black,urlcolor=black,citecolor=black}
\usepackage{natbib}

\usepackage[cmintegrals,cmbraces]{newtxmath}
\usepackage{ebgaramond-maths}

\makeatletter
\renewcommand{\maketitle}{
\twocolumn[
\begin{center}
\parbox{0.75\linewidth}{\centering\LARGE\@title\par}

\vspace{0.8em}

\parbox{0.4\linewidth}{
\centering
Henning Pohl \\
henning@cs.aau.dk\\
Aalborg University
}
\parbox{0.4\linewidth}{
\centering
Aske Mottelson \\
asmo@itu.dk\\
IT University Copenhagen
}
\end{center}
\vspace{2em}
]}
\makeatother

\author{Henning Pohl \& Aske Mottelson}
\title{Hafnia Hands: A Multi-Skin Hand Texture Resource for Virtual Reality Research}

\begin{document}

\maketitle

\section{Introduction}
We release a set of hand textures for use in virtual reality (VR) research.
The textures cover all six skin-tone levels of the Fitzpatrick scale~\citep{Fitzpatrick1988} in addition to three non-human variants.

Hand representations are an increasingly common way for users to interact in virtual environments and the hand appearance an important part of the user experience and bodily self-consciousness when immersed.
Our resource allows researchers to easily setup VR studies with visually realistic hand representations based on a standard skin tone scale.

Virtual hand representations are increasingly enabled by consumer-oriented VR technology through built-in hand tracking support (e.g., the Oculus Quest\footnote{\href{https://www.oculus.com/quest-2/}{https://www.oculus.com/quest-2/}}) as well as add-on products that can add hand tracking support to headsets that do not otherwise support it (e.g., the Leap Motion\footnote{\href{https://www.ultraleap.com/product/vr-developer-mount/}{https://www.ultraleap.com/product/vr-developer-mount/}}). 
While these systems usually provide hand models that work with their tracking, texturing of these models is limited.
For example, the Oculus and Leap Motion SDKs both do not include hand textures.

Unfortunately, creating detailed textures for existing hand models requires the expertise of a texture artist---a role not typically found in research teams.
There are textured hands available for purchase, but these usually use different underlying 3d models and hence are not directly compatible with the hand tracking data.
While it can be possible to adapt or create a hand skeleton that matches a model to that data, that in itself is a task that commonly requires expertise from a 3d artist.
For researchers that want to work with hand representations, but do not have expertise in 3d modeling, rigging, and texturing, there hence is a lack of easy to use options.

\subsection{Applications for Textured Hands in VR}
There are a range of VR research topics where hand textures are useful.
In particular, we see applications for the provided resource in (1) embodiment research and (2) conducting remote VR studies.

\subsubsection{Embodiment Research}

In VR, participants can experience illusory ownership of any imaginable body, making it an increasingly important research tool for understanding fundamental questions about the relationship between the body and the mind~\citep{Slater2009, Maister2015}.

Embodiment has been referred to participants feeling that, when given a virtual body, that body's ``\emph{properties are processed as if they were the properties of one's own biological body}''~\citep{Kilteni2012}.
One component of that is body ownership, the notion that a participant's ``\emph{body is the source of the experienced sensations}''~\citep{Kilteni2012}.
The relationship between the virtual body and the sense of body ownership is complex~\citep{Maselli2013}. 

Avatars, including hands, play an important role in embodiment research.
For example, more human-looking avatars do not necessarily lead to more body ownership~\citep{Lugrin2015}.
The sense of body ownership differs not just by visual realism, but also depending on other properties, such as perspective~\citep{Maselli2013}, or whether the hands are connected~\citep{Seinfeld2020}. 
Furthermore, VR embodiment research lacks standardized measures, tasks, or procedures, making it even more challenging to converge on theory-building.

The hand textures provided, can therefore aid three common challenges in VR embodiment research: 
(1) to match participants' real skin-tone to not conflict embodiment measures; 
(2) to enable research in the important of visual congruence (e.g., by mismatching participants' skin tone and virtual hand texture); and 
(3) to create a new standard procedure for employing humans hands in VR embodiment research.

\subsubsection{Remote VR Studies}
The infrastructure for remote VR studies is still developing and an active research area, and with the recent growing adoption of consumer-level VR devices, remote VR studies have now become a feasible alternative to lab studies \citep{Mottelson2021, Ratcliffe2021}.
Such crowdsourced VR experiments can replicate lab studies \citep{Ma2018} and enable more seamless recruitment of larger numbers and more diverse participants. 
Furthermore, remote studies can happen even at times when access to laboratories is restricted \citep{Steed2020}.

Before consumer VR devices were commonly available, previous studies have distributed \textit{Google Cardboard} kits to participants \citep{Mottelson2017, Steed2016}. 
Similarly, researchers have shown the feasibility of tapping into the exiting VR software ecosystem, for example, running studies inside the \textit{VRChat} application \citep{Saffo2020}.

Currently, the most popular VR devices for consumers are the Oculus Quest 1~and~2.
Both come with hand tracking support and thus allow for running a range of studies that benefit from this level of fidelity (e.g., studies of grasping, embodiment, or gesturing).
As remote studies commonly have a diverse set of participants, it is important that the hand representation they work with reflects this diversity.
Hence, the skin-tone variants we make available are an important component required for running these studies.

\subsection{VR Assets as Research Artifacts}
With our hand texture resource, we provide a piece of infrastructure for VR researchers.
While VR studies require a substantial amount of technical expertise to set up, resources like ours can ease that burden, but also allow for more more comparable results across studies.
In recent years, there has been a growing number of scientific resources similar to the Hafnia Hands.
For example, \citet{Regal2018} made assets available that make it easier to integrate questionnaires into VR scenes.
With \textit{NavWell}, \citet{Commins2020} released a tool for creating and running navigation experiments.
This standardizes the study format, but also substantially lowers the barriers for running such a study.
Most closely related to our hand texture resource is the \textit{Microsoft Rocketbox avatar library}~\citep{Gonzalez-Franco2020}, containing many rigged humanoid 3D models readily available as self-avatars, or as agents within VR environments.

\section{Method}
\begin{figure}[h!]
\includegraphics[width=\linewidth]{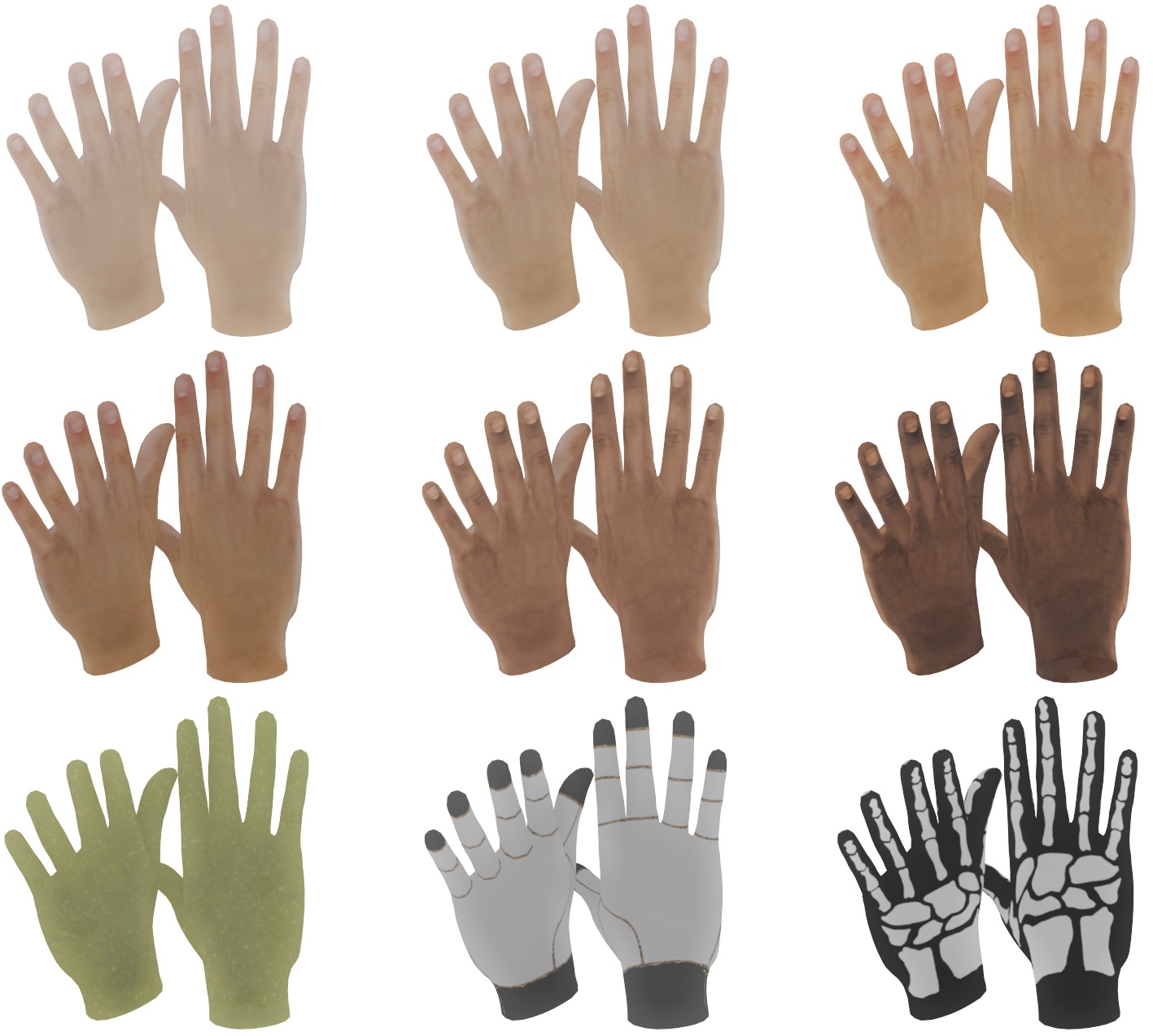}
\caption{We provide nine different hand textures: six skin-tone variants, an alien, a robot, and a skeleton hand. The skin-tone variants are designed to look realistic, while the non-human hands are more stylized.}
\label{fig:HandTextures}
\end{figure}

We commissioned the set of nine hand texture variants from a 3d artist on Fiverr.
Fiverr is a platform for connecting with freelancers, focused on creative tasks such as video editing or 3d modeling.
The set (shown in Figure~\ref{fig:HandTextures}) includes six different human skin-tones, as well as three non-human hands: a robot hand, an alien hand, and a skeleton hand.
Each variant comes with albedo, ambient occlusion, metallic, smoothness\slash{}roughness, and normal maps and hence is compatible with commonly used physically-based shading models (such as the \textit{Unity Standard Shader}).

The textures are designed to work directly with the default hand used by the Oculus Quest hand tracking.
Currently, this is the most widely used variant of hands for VR and also the one most suited for remote studies due to the large user base of consumers.
However, note that the textures can, in principle, be used with other hand models. 
Assuming the topology is similar (as should be the case for models of human hands) this would only require some changes to the UV coordinates of those models. 
Additionally, we provide hand textures compatible with the 3D hand models developed for the Oculus controllers; that way researchers can utilize our hand resource for studies using both free hand interaction and controller-based interaction.

We conducted an online and unsupervised study to evaluate the suitability of the hands for remote evaluations, and to get quantitative insights of their effects on embodiment. 
We evaluate the effects of skin-tone matching as well as owning the non-human hand variants.
The study, its hypotheses, method, procedure, and analyses approach was pre-registered\footnote{\href{https://osf.io/us39z/}{https://osf.io/us39z/}}.

We used a within-subjects design with hand texture as the only independent variable.
Hand texture had three levels: skin-tone matched, skin-tone mismatched, and non-human.
Participants completed three repetitions of a hand movement task with each hand texture, balanced using a latin square, for a total of nine trials per participant.

\subsection{Participants}
We recruited 112~participants from an internal email list of previous participants.
Participants were predominantly male (103 male, 6 female, and 3 participants who did not want disclose gender) and young, with an estimated mean age of 28.9 (SD 9.6).
The median reported skin tone was 3. Most participants reported medium skin tones, with 86\% in the 2--4 range on the Fitzpatrick scale (6$\times$ 1, 36$\times$ 2, 36$\times$ 3, 24$\times$ 4, 7$\times$ 5, and 3$\times$ 6 on the Fitzpatrick scale).
Participants conducted the study from 32~different countries, most commonly the USA (18), Poland (13), Germany (12), the UK (12), and Italy (8).
The participants were also experienced VR users, with the majority (51) having more than 100 hours of VR experience.

\subsection{Task}
We employed a novel study task where participants followed targets with their hands.
The task required participants to \textit{act with their hands}, while also forcing them to \textit{focus on their hands}.
Each trial took 60~seconds and was split in two equally long phases: one for each hand, ordered randomly.

During the task, participants had to keep their hand inside a framed box that appeared in front of them.
After appearing, the box moved along a randomly generated path within a $40 \times 60 \times 40$~cm volume.
We generated the path so each segment was at least 20~cm long.
Box movement slowed down at path nodes and accelerated in between, moving at an average overall speed of 0.6~m per second.
During each segment the box also performed a random rotation to between $-40$\textdegree and $90$\textdegree (left hand) $-90$\textdegree and $40$\textdegree (right hand) roll.
The boxes changed color to indicate whether participants' hand were contained within, prompting participants to follow study procedures.
When participants moved outside the boxes, we displayed a textual alert in the background asking them to return their hand inside.

\subsection{Procedure}
We first prompted participants for informed consent and collected demographic information, using a virtual questionnaire.
We then asked participants about their skin tone, which then formed the selection for the skin-tone matched condition.
For the skin-tone mismatched condition, we picked a hand texture three levels away from the participant's (i.e., $1 \rightarrow 4$, $3 \rightarrow 6$, $5 \rightarrow 2$, \ldots).
We always used the alien hand texture for the non-human condition.
Participants then completed nine trials (three with each hand) in random order.
Overall, the study took about 10~minutes.
A first warm-up round was used to determine tracking quality. 
If mean confidence for tracking was below 60\%, the participant would retry until quality was satisfactory, as per~\citep{Mottelson2021}.

\subsection{Measures}
For each trial we measured four dependent subjective variables using questionnaires displayed in VR:
To measure \textbf{Resemblance} we used the \textit{Features} question from \citep{Banakou2014}.
To measure \textbf{Agency} we used the \textit{Agency} question from \citep{Banakou2014}.
To measure \textbf{Body Ownership} we used the \textit{MyBody} question from \citep{Banakou2014}.
To measure \textbf{Humanness} we used a scale from \citep{Ho2017}, composed of five questions.

%
%
%

\begin{figure*}[h!]
\centering
\includegraphics[width=0.8\linewidth]{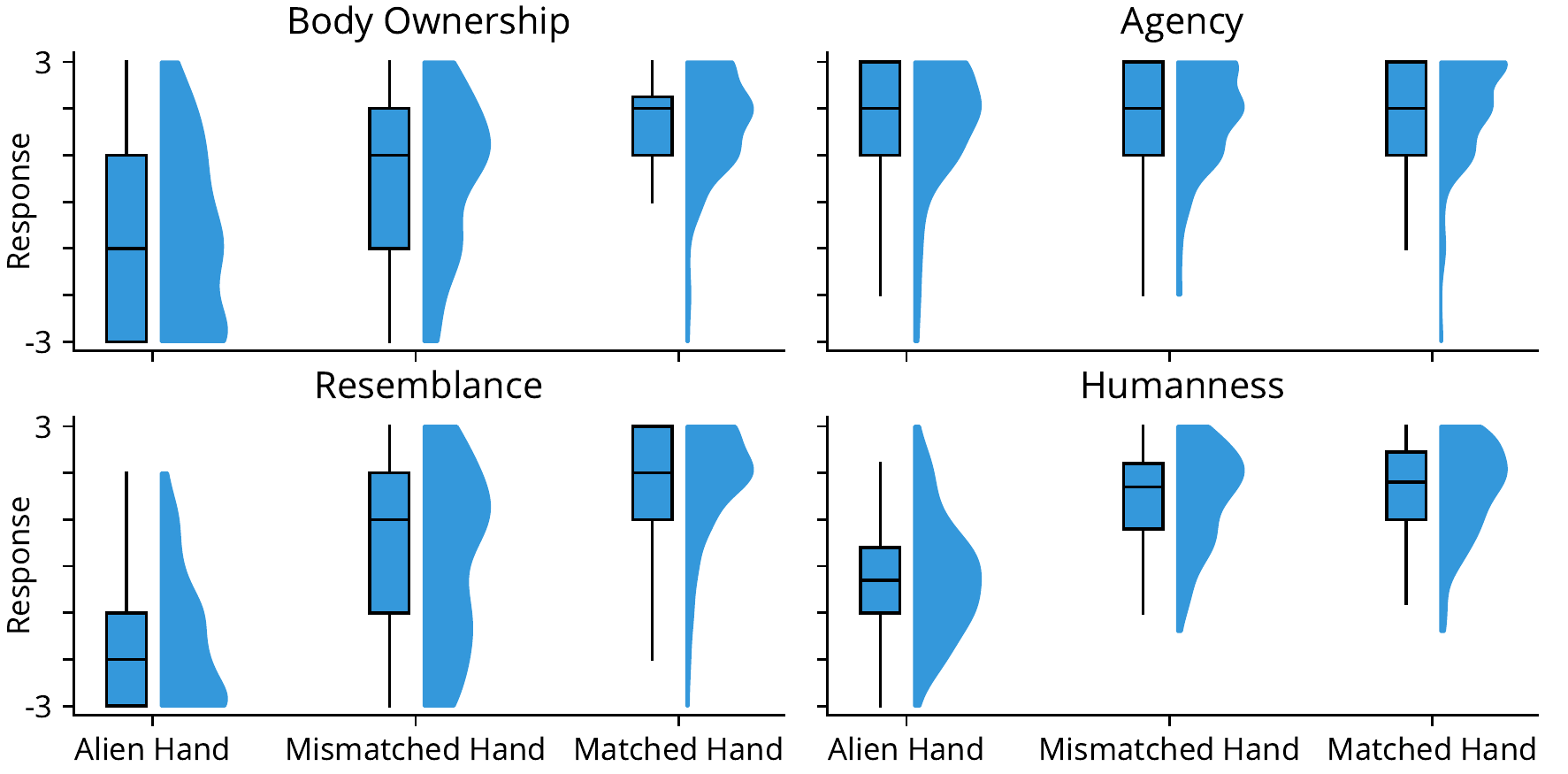}
\caption{Participants' responses to the four subjective measures we used for validating our hand texture dateset. Responses range from -3 (strongly disagree) to 3 (strongly agree), humanness shows aggregated scale based on five such responses.}
\label{fig:Validation}
\end{figure*}

\section{Validation}
From the participants responses we seek to validate the suitability of Hafnia hands for VR research.
In particular, when used for embodiment and remote studies, Hafnia hands need to (1) resemble participants' own hands and hence also elicit a higher sense of body ownership.
See Figure~\ref{fig:Validation} for an overview of how participants' rated their experience with the hands.

We test the effects of hand representation on body ownership, agency, and resemblance using Friedman tests for main effects.
For the humanness scale, we use a repeated-measure ANOVA.
We use Wilcoxon rank sum tests with Bonferroni correction for post-hoc comparisons.

We find an effect of hand representation on body ownership; $\chi^2(2) = 109.77, p < 0.001$.
All pairwise comparisons were significant ($p < 0.001$).
For agency, there was a significant main effect $\chi^2(2) = 26.27, p < 0.001$.
Yet, only the alien hand and matched hand were significantly different ($p = 0.04$).
We also found an effect of hand representation on resemblance; $\chi^2(2) = 150.31, p < 0.001$.
All pairwise comparisons were also significant ($p < 0.001$).
%
Finally, we also found a significant effect of hand representation on humanness; $F(2, 222) = 189.6, p < 0.001$.
Post-hoc tests found significant differences between the alien hand and the two realistic hands ($p < 0.001$).

In summary, the hands provided as part of Hafnia hands showed increased body ownership using a skin-tone matched texture; increased resemblance with a skin-tone matched texture; and decreased humanness with a alien texture. 
All registered hypotheses were hence found to be true.

These results also validate the suitability of Hafnia hands as a resource in embodiment and remote VR studies.
Particularly with respect to body ownership, skin-tone matched hands provide a benefit.
Furthermore, we validated that the provided non-human alien hand elicits the opposite response: low body ownership, visual resemblance, and humanness.
Hence, the set of provided hand textures covers a wide spectrum of subjective responses, which is required in a range of study designs.

\section{Limitations}
The sample for the evaluation was predominantly young and male, which has previously been shown to cause bias in VR research~\citep{Peck2020}. 
The hand textures visually resemble relatively young human hands, making the sample's age fit the study; however, future studies could entail textures matched and evaluated for seniors. 
The hand textures were, however, designed as gender neutral, whereas the evaluation consisted of predominantly males. 
This is a limitation of the evaluation, caused the bias of HMD owners' demographics.

As a manually-created resource, Hafnia hands are limited in the range of visual characteristics they cover.
In addition to lack of gender and age specificity, the hands also do not account for several other properties, such as: dirtiness, fingernail length\slash{}style\slash{}color, skin conditions, or finer nuances in skin-tone.
A potential solution for this are parametric hand texture models~\citep{Qian2020}.
However, to make use of the wide range of such models, a parameter acquisition step for each participant is required.
The quality of current model outputs also is not consistently at the same level as artist-created textures and does not include non-human textures at all.

While hands are perhaps the most important limb in embodiment research (e.g., the rubber hand illusion), future research resources and studies could consider creating resources enabling full body VR embodiment studies.

\section*{Data Availability Statement}
The hand textures released with this publication can be found in our Github repository\footnote{\url{https://github.com/henningpohl/Hafnia-Hands}}. 
The data for the evaluation is available at the same repository.

\section*{Ethics Statement}
Ethical review and approval was not required for the study on human participants in accordance with the local legislation and institutional requirements.
The participants provided their written informed consent to participate in this study.

\section*{Author Contributions}
All authors listed have made a substantial, direct and intellectual contribution to the work, and approved it for publication. 
Both HP and AM designed, implemented, and conducted the evaluation, HP conducted the statistical analyses; HP made figures, and HP and AM wrote the manuscript.

\section*{Conflict of Interest Statement}
The authors declare that the research was conducted in the absence of any commercial or financial relationships that could be construed as a potential conflict of interest.

\section*{Acknowledgments}
We'd like to thank Ranjeet Singh\footnote{\href{https://www.fiverr.com/google_jatt}{https://www.fiverr.com/google\_jatt}} for creating the hand textures.

\section*{Funding}
This research was supported by UCPH's Data+ pool under the agreement `Quantifying Body Ownership'.

\bibliographystyle{abbrvnat}
\bibliography{HandTextures}

\begin{thebibliography}{21}
\providecommand{\natexlab}[1]{#1}
\providecommand{\url}[1]{\texttt{#1}}
\expandafter\ifx\csname urlstyle\endcsname\relax
  \providecommand{\doi}[1]{doi: #1}\else
  \providecommand{\doi}{doi: \begingroup \urlstyle{rm}\Url}\fi

\bibitem[Banakou and Slater(2014)]{Banakou2014}
D.~Banakou and M.~Slater.
\newblock Body ownership causes illusory self-attribution of speaking and
  influences subsequent real peaking.
\newblock \emph{Proceedings of the National Academy of Sciences}, 111\penalty0
  (49):\penalty0 17678--17683, 2014.
\newblock ISSN 0027-8424.
\newblock \doi{10.1073/pnas.1414936111}.
\newblock URL \url{https://www.pnas.org/content/111/49/17678}.

\bibitem[Commins et~al.(2020)Commins, Duffin, Chaves, Leahy, Corcoran, Caffrey,
  Keenan, Finan, and Thornberry]{Commins2020}
S.~Commins, J.~Duffin, K.~Chaves, D.~Leahy, K.~Corcoran, M.~Caffrey, L.~Keenan,
  D.~Finan, and C.~Thornberry.
\newblock Navwell: A simplified virtual-reality platform for spatial navigation
  and memory experiments.
\newblock \emph{Behavior Research Methods}, 52\penalty0 (3):\penalty0
  1189--1207, Jun 2020.
\newblock ISSN 1554-3528.
\newblock \doi{10.3758/s13428-019-01310-5}.

\bibitem[Fitzpatrick(1988)]{Fitzpatrick1988}
T.~B. Fitzpatrick.
\newblock {The Validity and Practicality of Sun-Reactive Skin Types I Through
  VI}.
\newblock \emph{Archives of Dermatology}, 124\penalty0 (6):\penalty0 869--871,
  06 1988.
\newblock ISSN 0003-987X.
\newblock \doi{10.1001/archderm.1988.01670060015008}.

\bibitem[Gonzalez-Franco et~al.(2020)Gonzalez-Franco, Ofek, Pan, Antley, Steed,
  Spanlang, Maselli, Banakou, Pelechano, Orts-Escolano, Orvalho, Trutoiu,
  Wojcik, Sanchez-Vives, Bailenson, Slater, and Lanier]{Gonzalez-Franco2020}
M.~Gonzalez-Franco, E.~Ofek, Y.~Pan, A.~Antley, A.~Steed, B.~Spanlang,
  A.~Maselli, D.~Banakou, N.~Pelechano, S.~Orts-Escolano, V.~Orvalho,
  L.~Trutoiu, M.~Wojcik, M.~V. Sanchez-Vives, J.~Bailenson, M.~Slater, and
  J.~Lanier.
\newblock The rocketbox library and the utility of freely available rigged
  avatars.
\newblock \emph{Frontiers in Virtual Reality}, 1:\penalty0 20, 2020.
\newblock ISSN 2673-4192.
\newblock \doi{10.3389/frvir.2020.561558}.
\newblock URL
  \url{https://www.frontiersin.org/article/10.3389/frvir.2020.561558}.

\bibitem[Ho and MacDorman(2017)]{Ho2017}
C.-C. Ho and K.~F. MacDorman.
\newblock Measuring the uncanny valley effect.
\newblock \emph{International Journal of Social Robotics}, 9\penalty0
  (1):\penalty0 129--139, Jan 2017.
\newblock ISSN 1875-4805.
\newblock \doi{10.1007/s12369-016-0380-9}.

\bibitem[Kilteni et~al.(2012)Kilteni, Groten, and Slater]{Kilteni2012}
K.~Kilteni, R.~Groten, and M.~Slater.
\newblock The sense of embodiment in virtual reality.
\newblock \emph{Presence: Teleoperators and Virtual Environments}, 21\penalty0
  (4):\penalty0 373--387, 2012.
\newblock \doi{10.1162/PRES\_a\_00124}.

\bibitem[{Lugrin} et~al.(2015){Lugrin}, {Latt}, and {Latoschik}]{Lugrin2015}
J.~{Lugrin}, J.~{Latt}, and M.~E. {Latoschik}.
\newblock Avatar anthropomorphism and illusion of body ownership in vr.
\newblock In \emph{2015 IEEE Virtual Reality (VR)}, pages 229--230, 2015.
\newblock \doi{10.1109/VR.2015.7223379}.

\bibitem[Ma et~al.(2018)Ma, Cackett, Park, Chien, and Naaman]{Ma2018}
X.~Ma, M.~Cackett, L.~Park, E.~Chien, and M.~Naaman.
\newblock Web-based vr experiments powered by the crowd.
\newblock WWW '18, page 33–43, Republic and Canton of Geneva, CHE, 2018.
  International World Wide Web Conferences Steering Committee.
\newblock ISBN 9781450356398.
\newblock \doi{10.1145/3178876.3186034}.
\newblock URL \url{https://doi.org/10.1145/3178876.3186034}.

\bibitem[Maister et~al.(2015)Maister, Slater, Sanchez-Vives, and
  Tsakiris]{Maister2015}
L.~Maister, M.~Slater, M.~V. Sanchez-Vives, and M.~Tsakiris.
\newblock Changing bodies changes minds: owning another body affects social
  cognition.
\newblock \emph{Trends in Cognitive Sciences}, 19\penalty0 (1):\penalty0 6--12,
  2015.
\newblock ISSN 1364-6613.
\newblock \doi{https://doi.org/10.1016/j.tics.2014.11.001}.
\newblock URL
  \url{https://www.sciencedirect.com/science/article/pii/S1364661314002344}.

\bibitem[Maselli and Slater(2013)]{Maselli2013}
A.~Maselli and M.~Slater.
\newblock The building blocks of the full body ownership illusion.
\newblock \emph{Frontiers in Human Neuroscience}, 7:\penalty0 83, 2013.
\newblock ISSN 1662-5161.
\newblock \doi{10.3389/fnhum.2013.00083}.
\newblock URL
  \url{https://www.frontiersin.org/article/10.3389/fnhum.2013.00083}.

\bibitem[Mottelson and Hornb\ae{}k(2017)]{Mottelson2017}
A.~Mottelson and K.~Hornb\ae{}k.
\newblock Virtual reality studies outside the laboratory.
\newblock In \emph{Proceedings of the 23rd ACM Symposium on Virtual Reality
  Software and Technology}, VRST '17, New York, NY, USA, 2017. Association for
  Computing Machinery.
\newblock ISBN 9781450355483.
\newblock \doi{10.1145/3139131.3139141}.
\newblock URL \url{https://doi.org/10.1145/3139131.3139141}.

\bibitem[Mottelson et~al.(2021)Mottelson, Petersen, Lilija, and
  Makransky]{Mottelson2021}
A.~Mottelson, G.~B. Petersen, K.~Lilija, and G.~Makransky.
\newblock Conducting unsupervised virtual reality user studies online.
\newblock \emph{Frontiers in Virtual Reality}, 2021.
\newblock \doi{10.3389/frvir.2021.681482}.
\newblock URL
  \url{https://www.frontiersin.org/articles/10.3389/frvir.2021.681482/abstract}.

\bibitem[Peck et~al.(2020)Peck, Sockol, and Hancock]{Peck2020}
T.~C. Peck, L.~E. Sockol, and S.~M. Hancock.
\newblock Mind the gap: The underrepresentation of female participants and
  authors in virtual reality research.
\newblock \emph{IEEE Transactions on Visualization and Computer Graphics},
  26\penalty0 (5):\penalty0 1945--1954, 2020.
\newblock \doi{10.1109/TVCG.2020.2973498}.

\bibitem[Qian et~al.(2020)Qian, Wang, Mueller, Bernard, Golyanik, and
  Theobalt]{Qian2020}
N.~Qian, J.~Wang, F.~Mueller, F.~Bernard, V.~Golyanik, and C.~Theobalt.
\newblock {HTML: A Parametric Hand Texture Model for 3D Hand Reconstruction and
  Personalization}.
\newblock In \emph{Proceedings of the European Conference on Computer Vision
  (ECCV)}. Springer, 2020.

\bibitem[Ratcliffe et~al.(2021)Ratcliffe, Soave, Bryan-Kinns, Tokarchuk, and
  Farkhatdinov]{Ratcliffe2021}
J.~Ratcliffe, F.~Soave, N.~Bryan-Kinns, L.~Tokarchuk, and I.~Farkhatdinov.
\newblock Extended reality (xr) remote research: A survey of drawbacks and
  opportunities.
\newblock In \emph{Proceedings of the 2021 CHI Conference on Human Factors in
  Computing Systems}, CHI '21, New York, NY, USA, 2021. Association for
  Computing Machinery.
\newblock ISBN 9781450380966.
\newblock \doi{10.1145/3411764.3445170}.
\newblock URL \url{https://doi.org/10.1145/3411764.3445170}.

\bibitem[{Regal} et~al.(2018){Regal}, {Schatz}, {Schrammel}, and
  {Suette}]{Regal2018}
G.~{Regal}, R.~{Schatz}, J.~{Schrammel}, and S.~{Suette}.
\newblock Vrate: A unity3d asset for integrating subjective assessment
  questionnaires in virtual environments.
\newblock In \emph{2018 Tenth International Conference on Quality of Multimedia
  Experience (QoMEX)}, pages 1--3, 2018.
\newblock \doi{10.1109/QoMEX.2018.8463296}.

\bibitem[Saffo et~al.(2020)Saffo, Yildirim, Di~Bartolomeo, and
  Dunne]{Saffo2020}
D.~Saffo, C.~Yildirim, S.~Di~Bartolomeo, and C.~Dunne.
\newblock Crowdsourcing virtual reality experiments using vrchat.
\newblock CHI EA '20, page 1–8, New York, NY, USA, 2020. Association for
  Computing Machinery.
\newblock ISBN 9781450368193.
\newblock \doi{10.1145/3334480.3382829}.
\newblock URL \url{https://doi.org/10.1145/3334480.3382829}.

\bibitem[Seinfeld and Müller(2020)]{Seinfeld2020}
S.~Seinfeld and J.~Müller.
\newblock Impact of visuomotor feedback on the embodiment of virtual hands
  detached from the body.
\newblock \emph{Scientific Reports}, 10\penalty0 (1):\penalty0 22427, Dec 2020.
\newblock ISSN 2045-2322.
\newblock \doi{10.1038/s41598-020-79255-5}.

\bibitem[Slater et~al.(2009)Slater, Pérez~Marcos, Ehrsson, and
  Sanchez-Vives]{Slater2009}
M.~Slater, D.~Pérez~Marcos, H.~Ehrsson, and M.~Sanchez-Vives.
\newblock Inducing illusory ownership of a virtual body.
\newblock \emph{Frontiers in Neuroscience}, 3:\penalty0 Article 29, 2009.
\newblock ISSN 1662-453X.
\newblock \doi{10.3389/neuro.01.029.2009}.
\newblock URL
  \url{https://www.frontiersin.org/article/10.3389/neuro.01.029.2009}.

\bibitem[Steed et~al.(2016)Steed, Frlston, Lopez, Drummond, Pan, and
  Swapp]{Steed2016}
A.~Steed, S.~Frlston, M.~M. Lopez, J.~Drummond, Y.~Pan, and D.~Swapp.
\newblock An ‘in the wild’ experiment on presence and embodiment using
  consumer virtual reality equipment.
\newblock \emph{IEEE Transactions on Visualization and Computer Graphics},
  22\penalty0 (4):\penalty0 1406--1414, 2016.
\newblock \doi{10.1109/TVCG.2016.2518135}.

\bibitem[Steed et~al.(2020)Steed, Ortega, Williams, Kruijff, Stuerzlinger,
  Batmaz, Won, Rosenberg, Simeone, and Hayes]{Steed2020}
A.~Steed, F.~R. Ortega, A.~S. Williams, E.~Kruijff, W.~Stuerzlinger, A.~U.
  Batmaz, A.~S. Won, E.~S. Rosenberg, A.~L. Simeone, and A.~Hayes.
\newblock Evaluating immersive experiences during covid-19 and beyond.
\newblock \emph{Interactions}, 27\penalty0 (4):\penalty0 62–67, July 2020.
\newblock ISSN 1072-5520.
\newblock \doi{10.1145/3406098}.
\newblock URL \url{https://doi.org/10.1145/3406098}.

\end{thebibliography}

\end{document}